\documentclass[%
 reprint,
 amsmath,amssymb,
 aps 
]{revtex4-2}

\usepackage{graphicx}
\usepackage{dcolumn}
\usepackage{bm}

\begin{document}

\preprint{APS/123-QED}

\title{Parametric All-Optical Modulation on Chip}
\author{Zhan Li $^{1,2}$, Jiayang Chen $^{1,2}$, Zhaohui Ma $^{1,2}$, 
Chao Tang $^{1,2}$, Yong Meng Sua $^{1,2}$, Yu-Ping Huang $^{1,2,3,\dag}$}
\affiliation{$^1$Physics Department, Stevens Institute of Technology, Hoboken, New Jersey, 07030, United States \\
$^2$Center for Quantum Science and Engineering, Stevens Institute of Technology, Hoboken, New Jersey, 07030, United States\\ 
$^3$Quantum Computing Inc, Hoboken, NJ, 07030, United States \\
$^\dag$yuping.huang@stevens.edu}


\begin{abstract}
We demonstrate parametric all-optical modulation in a periodically-poled lithium niobate microring resonator on chip. It employs quantum Zeno blockade between two distinct waves, a signal and a pump, through their sum-frequency generation at a large per-photon efficiency of $8.2$ MHz. With nanosecond pump pulses at $6$ mW peak power, $85.7\%$ modulation extinction is observed, achieving an efficiency improvement of over 30 times compared to previous implementations. With only $2$ mW pump peak power, $43.0\%$ modulation extinction is observed for a doubly-stronger signal at $4$ mW. This demonstrates, for the first time, that optical transistors with cascadability and fan-out are possible with just parametric nonlinear optics. These results, together with inherent advantages in such photonic integrated circuits, open the door to scalable technology for all-optical and quantum information processing. 
\end{abstract}

\maketitle


\noindent{\it Introduction---} Quantum Zeno effect occurs when a coherently-evolving system is coupled strongly to external degrees of freedom, with the result that the evolution is suppressed or frozen \cite{misra1977zeno,peres1980zeno,joos1984continuous,ai2010quantum}. This counter-intuitive and intriguing quantum phenomenon is implied in the postulates of quantum mechanics, where performing measurement on a quantum object collapses its wavefunction onto an eigenstate of the measuring apparatus. Here, the strong coupling effectively implements fast repeated measurements, thereby forcing the system to remain in the initial state. Since its first demonstration in trapped ions in 1990 \cite{itano1990quantum}, the Zeno effect has been widely studied \cite{elitzur1993quantum,kwiat1995interaction,kwiat1996quantum,tsegaye1998efficient,kwiat1999high}, and explored for exotic applications in quantum communications \cite{cao2017direct}, quantum computing \cite{hosten2006counterfactual}, all-optical information processing \cite{huang2011interaction,strekalov2014progress,chang2014quantum}, and so on. Among them, quantum Zeno blockade (QZB) was analyzed and demonstrated in nonlinear optical systems, including those of $\chi^{(2)}$ waveguides \cite{huang2010interaction,liu2023engineering}, etalons with sandwiched crystals \cite{mccusker2013experimental}, whispering-gallary mode cavities \cite{strekalov2014progress}, a thin-film microdisk \cite{chen2017observation}, and aluminum nitride microrings \cite{guo2018all}. 

Optical QZB occurs in an open quantum system consisting of certain optical modes coupled with external degrees of freedom through a nonlinear-optical channel. When the channel is ``fast'' (i.e., the nonlinear interaction is strong), the occupation of a mode can transiently alter the system condition and suppress its coherent dynamics, including “block” additional photons from coupling into a certain mode \cite{jacobs2009all,huang2010interactionf,huang2011interaction,huang2012antibunched}. Here, the mode can correspond to a standing wave in an optical cavity, or a traveling wave in a waveguide or optical fiber. The coherent dynamics can be the coupling from a bus waveguide to the cavity, or evanescent coupling between two waveguides. In a cavity setting, QZB can be implemented through sum-frequency generation (SFG) or difference-frequency generation (DFG), where populating the cavity with pump photons frustrates the cavity resonance condition for signal photons at a different wavelength \cite{huang2010interactionf,huang2011interaction,mccusker2013experimental,strekalov2014progress}. As a result, the signal is deflected from the cavity, and the frequency generation does not occur in the asymptotic limit. During this process, the pump photons act effectively as a probe dynamically monitoring if signal photons have entered the cavity, by coupling the latter to another lossy mode \cite{sun2013photonic}. When the monitoring is fast, the Zeno effect prevents the signal from the cavity, thus the blockade effect \cite{misra1977zeno,peres1980zeno,joos1984continuous,itano1990quantum}. 

While QZB is a quantum Zeno effect, when the signal is a coherent light beam, it can be intuitively understood in a classical-optics picture. Let's consider the case of a signal resonantly coupling to a cavity through a bus waveguide. When there is no pump, the signal couples into the cavity by slowly building up its field amplitude inside the cavity. Concurrent with this process, a small portion of the cavity field leaks out to the bus waveguide and destructively interferes with the signal remaining in the waveguide. It is this destructive interference that will cause the majority of the signal to enter the cavity. However, when there is a pump in the cavity to induce strong SFG or DFG, any signal in the cavity will be quickly converted to a different wavelength and effectively lost. This will prohibit its cavity-field building up and thus disrupt the destructive interference. As a result, the majority of the signal will continue to propagate in the bus waveguide and not enter the cavity \cite{huang2010interactionf}. 

Hence, QZB provides a counter-intuitive implementation of the interaction between optical signals: one to control another without them physically overlapping to interact \cite{kwiat1995interaction,kwiat1996quantum}. Furthermore, unlike the photon blockade mediated by single emitters, here no excitation of any kind is created, but the interaction occurs due to the mere potential for a parametric nonlinear optical process (which asymptotically does not occur) \cite{huang2010interaction,huang2010interactionf}. These distinct interaction-free and excitation-free features together eliminate the occurrence of phase noise, photon dissipation, and quantum state decoherence (e.g., via spontaneous emission) \cite{shapiro2006single}. It thus promises to overcome the fundamental challenges facing all-optical processing of quantum signals, including realizing deterministic interaction between single photons for scalable photonic quantum computing \cite{sun2013photonic}. The same interaction-free and excitation-free implementation of interaction makes QZB a strong candidate for optical transistors \cite{miller2010optical}, the backbone technology for of the future all-optical information processing. Asides from the above practical advantages in noise, loss, and decoherence, QZB naturally support superior cascadability, fan out, and absence of critical biasing, as critical for practical optical logic. This is because QZB effect transiently changes the cavity condition, so that the blockade effect is independent of the signal's power level and detailed waveforms. It thus allows a weaker pump to control a strong signal, or the output of one stage to drive multiple subsequent stages. Notably, as QZB only utilizes nonlinear optics, there is no energy conversion from optics to electronics or atomic, so that the otherwise detrimental heat sinking or electrical interference issues are eliminated. 

In view of those potentials, here we demonstrate, for the first time, a chip-integrated device for QZB and efficient all-optical modulation with cascadability and fan-out. Unlike previous QZB demonstrations, we use etched and encapsulated microring cavities on thin-film lithium niobate \cite{zhu2021integrated,lin2020advances}. By strong cavity confinement and quasi-phase matching, we observe high modulation extinction ratio (MER) requiring only milliwatt peak power, pico-joule pulse energy, corresponding to over orders of magnitude improvement over the state of the art (see Table \ref{tab:results compare} below). For cascadability and fan out, we also demonstrate that a weak pump can modulate a twice as strong signal. Together, our results highlight QZB in integrated lithium niobate nanophotonics as an enabling candidate for the future all-optical logic, photonic quantum computing, and beyond.  
\begin{table}[htbp]
\centering

\begin{tabular}{ccc}
\hline
Reference \; &  Pump Peak Power \;  &  MER \\
\hline
\cite{mccusker2013experimental} &  $17$ W& $\sim 80\%$ \\
 \cite{chen2017observation} &  $680$ mW& $\sim 51\%$ \\
 \cite{guo2018all} &  $180$ mW& $\sim 80\%$ \\
 This work & $6$ mW& $85.7\%$ \\
\hline
\end{tabular}
\caption{All-optical Zeno Effect Demonstrations. The MER of other works are derived from their modulation results.}
  \label{tab:results compare}
\end{table}

\noindent{\it Experiments and simulations} The periodically poled lithium niobate micro-ring (PPLNMR), whose input-output is through a pulley bus waveguide, was fabricated and characterized using procedures detailed in our previous works \cite{chen2019ultra,ma2020ultrabright,chen2021photon,ma2023highly}. We utilize the largest tensor $d_{33}$ of lithium niobate's $\chi^{(2)}$ in 600 nm z-cut samples and fabricate PPLNMR chip with a photon-photon coupling coefficient of $g=8.2$ MHz (angular frequency) and an overall intrinsic Quality factor $Q_i=7\times10^5$. It supports quasi-phase matched sum-frequency generation among three quasi-transverse magnetic (quasi-TM) modes, the signal at $\lambda_s=1545.73$ nm, the pump at $\lambda_p=1558.01$ nm, and their sum-frequency at $\lambda_f=775.93$ nm. Their intrinsic decay rates due to material absorption and scattering losses, external decay rates due to coupling with the pulley waveguide, and the total rates, are measured and listed in Table \ref{tab:measured Q factor}. As shown, for all three waves, the external decay is faster than the intrinsic decay by at least four times, which puts this chip in the significant over-coupling regime, as needed for most optical information processing. In this experiment, the pump acts as the control to frustrate the cavity resonance and modulate the signal, and the sum-frequency light is nearly not generated when the pump is strong. 
\begin{table}
\centering

\begin{tabular}{cccc}
\hline
 & Intrinsic Decay & External Decay & Total Decay \\
\hline
Pump & $1.6$ GHz & $8.45$ GHz & $10.1$ GHz \\
Signal & $1.8$ GHz & $7.6$ GHz & $9.4$ GHz \\
Sum-frequency & $2.0$ GHz & $12.1$ GHz  & $14.1$ GHz \\
\hline
\end{tabular}
\caption{Decay rates of Pump mode, Signal mode and Sum-frequency mode in angular frequency units. The intrinsic decay rates are derived by intrinsic Q factor, which is $7\times10^5$ at 1550 nm, corresponding to TM mode propagation loss of 40 dB/m}
  \label{tab:measured Q factor}
\end{table}
Note that those measurements in Table \ref{tab:measured Q factor} were performed at very low light intensity level, when the cavity is ``cold'' (room temperature). During experiment, the cavity may be heated up under high light intensity, and those parameters may change modestly. Meanwhile, the phase matching condition will change, which will require small change in chip temperature and the Pump and Signal wavelength to restore. 

The cavity dynamics can be modeled by the following equations \cite{chin1998design}:
\begin{eqnarray}
& &\frac{dC_p}{dt}    =(i\delta_p-\frac{\kappa_{t,p}}{2})C_p-ig^*C^*_sC_{f}-i\sqrt{\kappa_{e,p}}F_p \label{eq1} \\
& &\frac{dC_s}{dt}    =(i\delta_s-\frac{\kappa_{t,s}}{2})C_s-ig^*C^*_pC_{f}-i\sqrt{\kappa_{e,s}}F_s \label{eq2}\\ 
& &\frac{dC_{f}}{dt} =(i\delta_{f}-\frac{\kappa_{t,f}}{2})C_{f}-igC_pC_s \label{eq3}\\
& &O_p  =i\sqrt{\kappa_{e,p}}C_p+F_p\label{eq4} \\
& &O_s  =i\sqrt{\kappa_{e,s}}C_s+F_s\label{eq5} \\
& &O_f  =i\sqrt{\kappa_{e,f}}C_f\label{eq6}
 \end{eqnarray}
\noindent In the above equations, the subscripts $\sigma=p,s,f$ denote the pump, signal, and sum-frequency waves, respectively. $F_\sigma$ is the external wave at the input. $C_{\sigma}$ is the amplitude of the cavity field normalized such that $|C_{\sigma}|^2$ gives the photon number in the cavity. $O_{\sigma}$ is the cavity output. The total cavity decay rate $\kappa_{t,\sigma}$ is the sum of the intrinsic decay $\kappa_{i,\sigma}$ due to cavity losses and the external decay $\kappa_{e,\sigma}$ through coupling with the bus waveguide. $\delta_{\sigma}$ is the laser-cavity detuning, which is zero under ideal phase matching. With cold cavity $\kappa_{\sigma}$, the simulation results are calculated numerically.

\begin{figure*}
\centering
\includegraphics[width=0.9\linewidth]{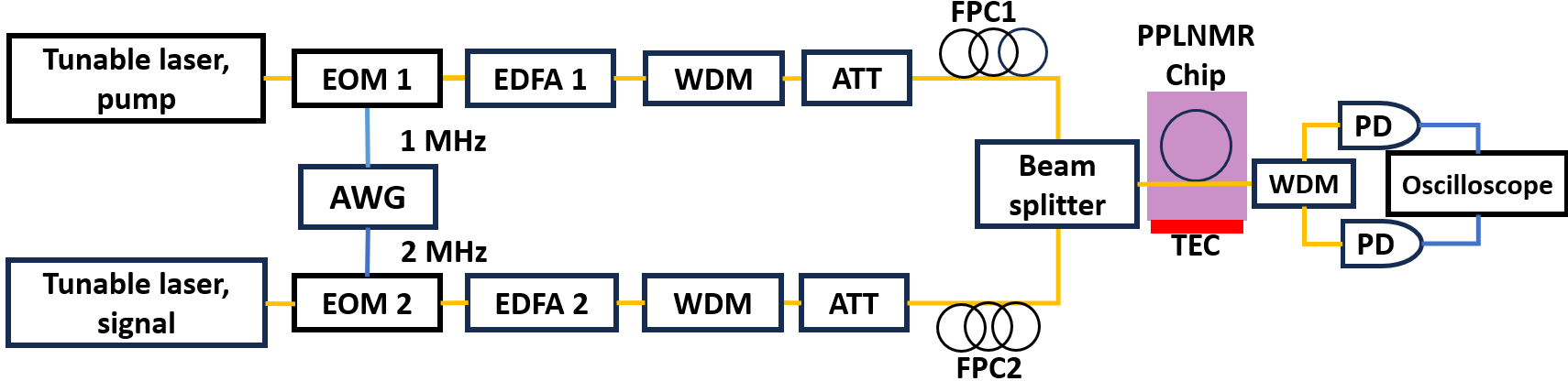}
\caption{Experiment schematic. Orange lines indicate optical connections while blue lines are for electronic connections. Lensed fibers are used for coupling lights in and out chip. TEC lies under the chip for temperature controlling.}
\label{fig:setup}
\end{figure*}


The experiment setup is depicted in the Fig.~\ref{fig:setup}. It consists of two narrow-band tunable lasers (Santec TSL-550/570), each followed by electronics and optics for generating synchronized optical pulse. They include an arbitrary waveform generator (AWG) generating synchronous electrical pulses to drive electric-optical modulators (EOMs) at $1$ MHz for the pump light and $2$ MHz for the signal light. The resulting optical pulse trains each undergo amplification by an Erbium-doped fiber amplifier (EDFA), followed by wavelength division multiplexers (WDMs) to filter out excessive ASE noise. Then, attenuators (ATTs) are used to vary each optical power level, before combining via a beamsplitter and directed jointly into the PPLNMR chip via a lensed fiber. The chip sits on a thermoelectric cooler (TEC) that controls and stabilizes its temperature with $0.01 C^\circ$  precision. After the chip, another WDM is used to separate the output pump and signal, before each is detected by a photodetector (R2860D receiver module, 10 GHz) and an oscilloscope (HP Agilent 54750a, 20 GHz). The waveguide propagation loss is approximately 40 dB/m, as inferred from our intrinsic Q factor of $7\times10^5$. The bus waveguide is 4 mm long, which gives ignorable device insertion loss. For the lensed fiber to chip edge coupling, the loss mainly comes from the mode mismatching between fibers and the waveguide, which is 5 dB per facet.\\

\noindent{\it Continuous-wave Signal and Pulsed Pump---} We first demonstrate all-optical modulation where a continuous-wave (CW) signal is modulated by a pulsed pump (as the probe for QZB). To minimize thermal and photo-refractive (PR) effects \cite{chen2020efficient}\cite{surya2021stable}\cite{shams2022reduced}, the signal is prepared in a rectangular waveform with $10$-ns quasi-Continuous Wave (quasi-CW) pulses at $2$ MHz repetition rate and $0.2$ mW peak power on chip (note that the power number to be quoted for the rest of the paper always refers to the peak power in bus waveguide before the cavity). The pump pulses, on the other hand, are in Gaussian form with $2$-ns full width half maximum (FWHM), at $1$ MHz repetition rate, and are set in the middle of the signal pulses. 

\begin{figure*}
\centering
\includegraphics[width=0.98\linewidth]{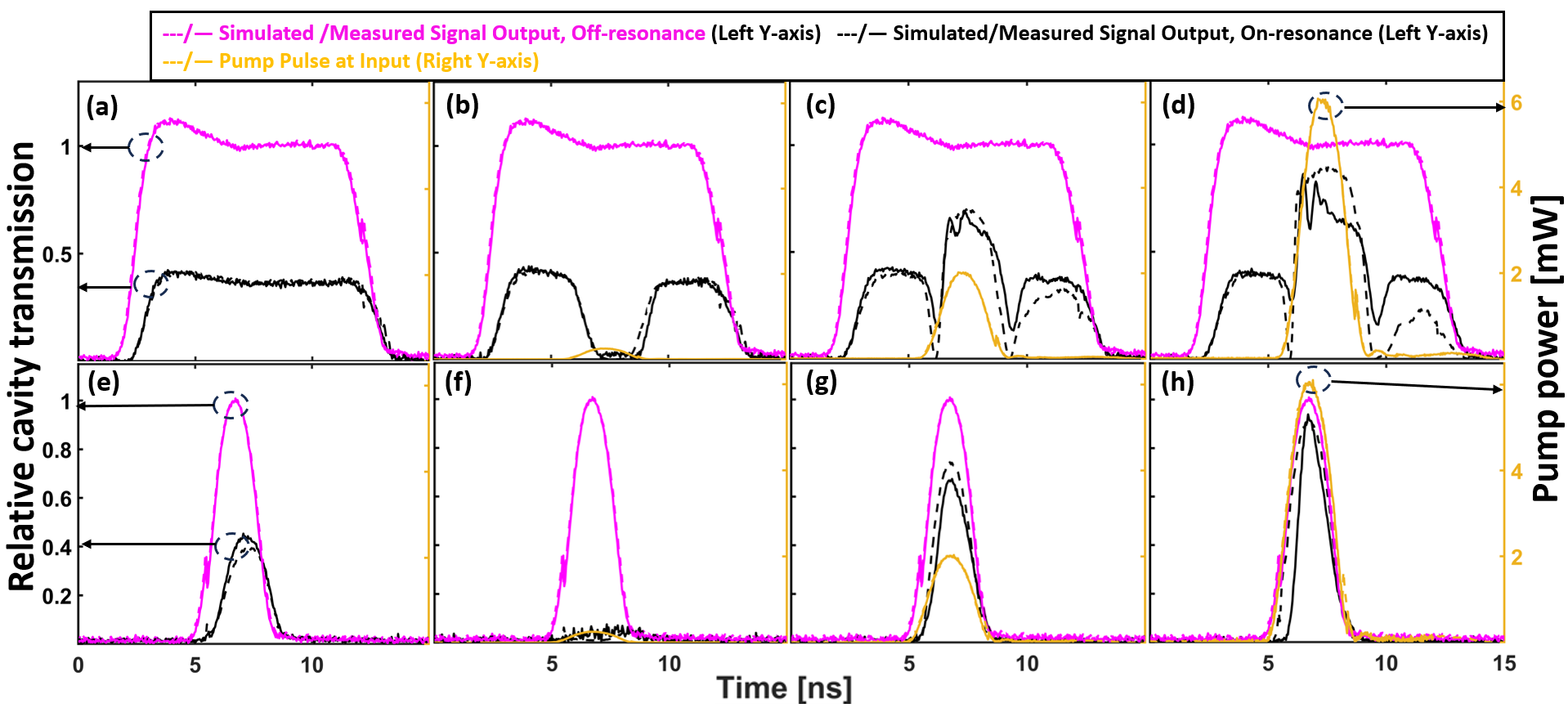}
\caption{Measured and simulated cavity transmission under various power settings. The top panels plot the results for the case of quasi-CW signal and pulsed pump, with the signal transmission under different conditions: (a) without pump, (b) pump peak power at $0.25$ mW, (c) pump peak power at $2$ mW, and (d) pump peak power at $6$ mW. The bottom panels plot the signal transmission in the case of pulsed signal and pulsed pump, where, similar to the top panels, (e) is the result without pump, and (f), (g), (h) are with pump peak power at $0.25$ mW, $2$ mW, and $6$ mW, respectively. In all figures, the solid curves represent the experimental results. The dashed curves are the simulation results. The pink curves are the signal relative transmission under off-resonance. The black curves are the signal relative transmission on resonance with different pump peak powers. The orange curves give the pump pulse profiles at the input.}
\label{fig:weak_sig_fit}
\end{figure*}

The results are shown in Fig.~\ref{fig:weak_sig_fit}(a-d). When there is no pump in cavity, the on-resonance signal transmission loss through the chip is $63.5\%$ as compared to the off resonance case, due to the signal mode over-coupling. Then, as shown in Fig.~\ref{fig:weak_sig_fit}(b), when pump pulses with low optical power ($0.25$ mW) are applied, a dip is created in their overlapping regime, with the transmission dropping to nearly zero (e.g., $\sim$ 100\% loss). This is because the pump drives SFG in the cavity, inducing additional losses and causing the cavity to transition from overcoupling (in the absence of the pump) to critical coupling for the signal. 

In Figure.~\ref{fig:weak_sig_fit}(c), as the pump power gradually increases to $2$ mW, the signal after cavity transmission bounces back. At the peak of the pump pulses, the signal mode transmission loss is reduced to $32.2\%$, which is about half of the on-resonance case. This is the result of quantum Zeno blockade: when the SFG-induced cavity loss is larger than the coupling decay, the signal is effectively under coupled from the cavity, so that it mostly does not enter the cavity \cite{huang2010interaction,mccusker2013experimental}. The MER, calculated as the percentage reduction of transmission due to the pump in cavity, is $49.3\%$. As shown in Fig.~\ref{fig:weak_sig_fit}(d), further increasing the pump power to $6$ mW improves the MER to $74.5\%$ with even lower transmission loss $14.8\%$. The trenches flanking the peak indicates effective critical coupling for the signal in those regimes, where the pump power induced cavity loses for the signal through SFG but not strong enough to cause QZB. In both Figs.~\ref{fig:weak_sig_fit}(c) and \ref{fig:weak_sig_fit}(d), the SFG at the pump peak is suppressed compared to the weaker pump case in Fig.~\ref{fig:weak_sig_fit}(b). Again, this is because QZB blocks the signal from entering the cavity.    

Also plotted in Figs.~\ref{fig:weak_sig_fit}(a-d) are the simulation results by solving Eqs.(\ref{eq1}-\ref{eq6}) with only the separately measured cold-cavity parameters in Table \ref{tab:measured Q factor}, without using any fitting parameter. Here, the signal output power is 73 $\mu$W when it is on resonance and there is no pump, which matches perfectly with the simulation result. When the pump is 0.25 mW, the measured signal output power is reduced to 5 $\mu$W (versus 0.3 $\mu$W as simulated). When the pump is 2 mW, the signal output power is measured to be 136 $\mu$W, while the simulation gives 146 $\mu$W. Further increasing the pump to 6 mW, it becomes 170 $\mu$W as measured and 180$\mu$W as simulated. As shown, the theoretical results follow closely the experimental results, indicating a good modeling of underlying dynamics. The disagreement in Figs.~\ref{fig:weak_sig_fit}(c) and (d) may due to the an over-estimate of the SFG efficiency $g$ and slight shift of the signal resonance peak in a ``hot'' cavity with high pump power. Note that in Figs.~\ref{fig:weak_sig_fit}(c-d), the transmitted signals have some oscillations around the peaks. This could be due to the actual coupling condition for the SF wave is less over-coupled, so that a strong SF field is trapped in the cavity, causing transient parametric oscillation. More discussions can be found in the Supplemental Materials.   \\

\noindent{\it Pulsed Signal and Pulsed Pump---} Next we demonstrate the modulation of signal pulses by pump pulses (the latter acting as the probe for QZB). The results for Gaussian signal pulses at $2$-ns FWHM and with the same peak power are shown in Fig.\ref{fig:weak_sig_fit}(e-h). With a much lower average power for the signal, the chip temperature and thus the cavity-coupling condition are changed, giving a $53.9\%$ on-resonance transmission loss (which indicates higher over-coupling). As seen, with the same pump applied, the modulation results follow the trend with the above quasi-CW case. That's when the signal overlapped with 0.25-mW pump pulses, the signal undergoes further suppression to critical coupling. Increasing the pump power to $6$ mW results in the signal transmission loss lowered to $7.6\%$ and MER reaching $85.7\%$. 

The simulation results plotted in Fig.\ref{fig:weak_sig_fit}(e-h), using the measured pulse profiles and amplitudes and cold-cavity parameters in Table \ref{tab:measured Q factor}, give a signal transmission dynamic mostly aligned with the experiment results. Here, without the pump, the signal peak power after modulation is measured to be 92 $\mu$W versus 73 $\mu$W as simulated. When the pump is 0.25 mW, the measured and simulated power values at the output are 9 $\mu$W and 1 $\mu$W, respectively.  They increase to 132 $\mu$W and 146 $\mu$W for pump power at 2 mW, and 185 $\mu$W and 182 $\mu$W for 6 mW pump power.


\begin{figure}[ht]
\centering
\includegraphics[width=\linewidth]{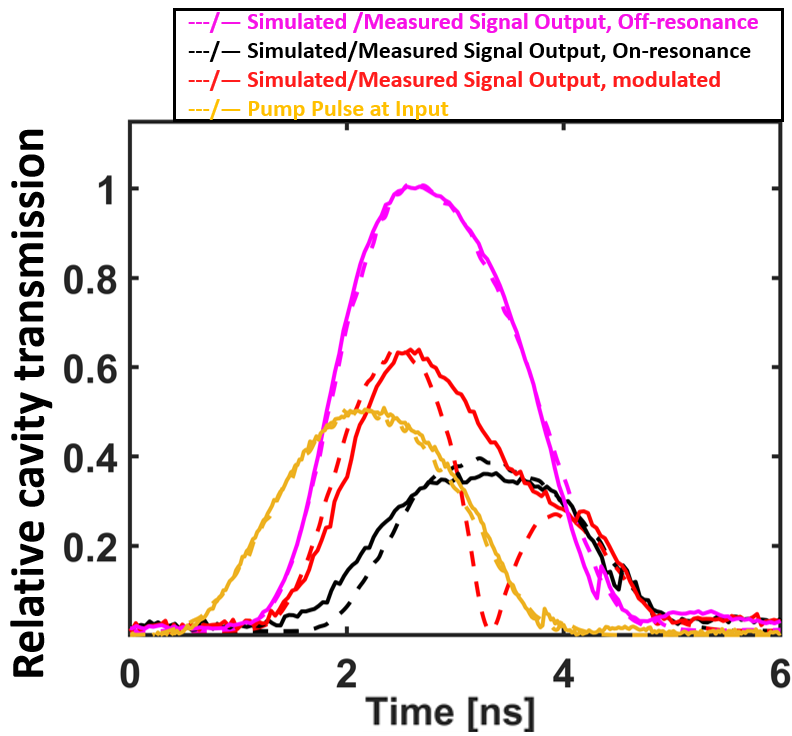}
\caption{Cavity transmission for the case of strong signal and weak pump. Solid and dashed curves represent experimental and simulation results, respectively. Black and pink curves are the transmitted pulse profiles of the signal when it is on and off resonance, respectively. Orange curves are the transmitted pump pulse profiles. Red curves show the transmitted signal pulse profile modulated by the pump.}
\label{fig:strong_s_fit}
\end{figure}

\begin{figure}[ht]
\centering
\includegraphics[width=1\linewidth]{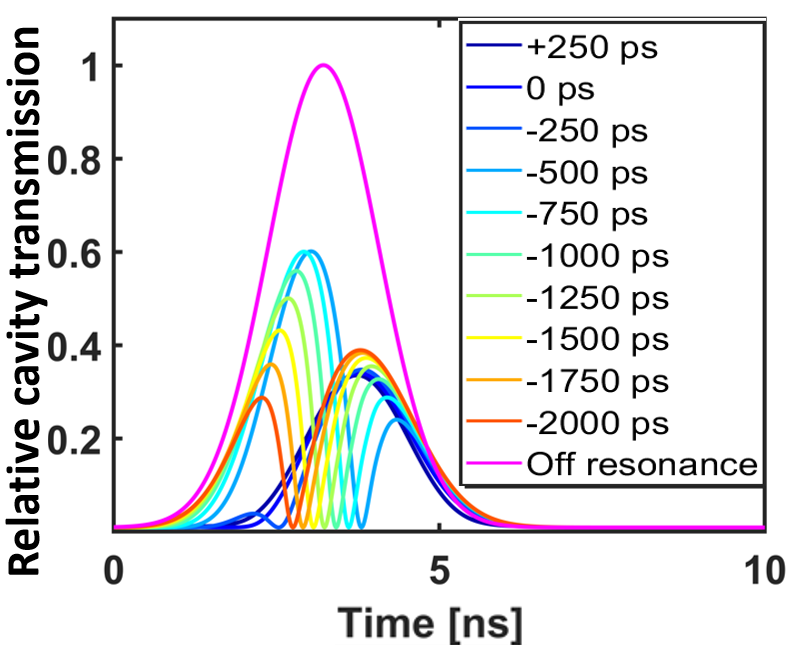}
\caption{Simulated modulation effects as a function  of the signal arrival time relative to the pump, where ``+'' and ``-'' in the legend indicate the signal arriving before and after the pump, respectively. }
\label{fig:strong_s_delay}
\end{figure}

\noindent{\it  Strong Signal and Weak Pump---} When the power of pump pulses reaches the threshold to frustrate the cavity, they can modulate and switch signals of much higher power. This unique feature, originating from QZB, is essential for scalability, fan out, and absence of critical bias in practical optical transistors. In Figure.~\ref{fig:strong_s_fit}, we demonstrate weaker pump pulses modulating stronger signal pulses. The signal and pump pulse shapes are the same with Fig.~\ref{fig:weak_sig_fit}, but with the pump at $2$ mW and the signal at $4$ mW.  As shown, the signal on resonance transmission loss is $63.7\%$. By adjusting the phase on AWG, the signal arrival time is tuned to achieve the best MER of $43.3\%$ at a $600$ ps delay from the pump. Its transmission loss is reduced to $36.1\%$. For a better understanding of the modulation as a function of the relative delay time, we simulate the signal transmission as the relative arrival time is swept. The results are plotted in Fig.~\ref{fig:strong_s_delay}, where the best MER of 35.5\% is obtained when the signal arrives 500 ps after the pump. More results can be found in the Supplemental Materials.

Figure.~\ref{fig:strong_s_fit} also plots the simulation results by using measured pulse profiles, amplitudes, and the cold-cavity parameters, without any fitting parameters. As seen, they are in good agreement with the experimental results. The on-resonance transmission loss is $60.5\%$ and the MER is $35.5\%$, with $39.9\%$ transmission loss under the pump modulation. Compared with the experiment results, the simulated MER is slightly lower. This is attributed to that the actual signal coupling is weaker than the measured cold-cavity case. Also, the simulated signal transmission shows a ringing bell oscillation around 0.5 ns duration. This is because at the pump pulse tail, SFG is not strong enough to prohibit the signal from entering the cavity via QZB, but rather causing critical coupling for the signal. Further into the tail, the SFG is diminished, because by then the pump has exited the cavity. Thus the signal ``sees'' an overcoupling cavity, so that its transmission loss is less. Yet, with the detector fast enough, this ringing is not observed in experiment, which indicates other effects, such as those of photorefractive (PR), thermo-optic, and second-harmonic generation (SHG), may need to be added to the current model for a more accurate description of the system dynamics.  This will remain a task for the future studies.  

\noindent{\it Conclusion and Discussion---} We have demonstrated parametric all-optical modulation with high efficiency and presented clear evidences towards its uses for practical optical transistor technology. It employs quantum Zeno blockade in a periodic poled microring in thin-film lithium niobate, offering rich and counter-intuitive nonlinear optical phenomena. Our simulation results agree well with the experimental results, with subtle differences coming from side effects not captured by our model, including those of PR, thermo-optic, and SHG. They call for suppression of those effects for improved performance, e.g., by addressing PR by MgO$_2$ doping and addressing SHG by narrowing phase matching. To reduce the thermo-optic effect, one can continuously monitor the signal at an adjacent period to make sure it stays on resonance and fine adjustment of the phase matching condition by tuning the tunable lasers and the TEC.

From our results, the modulation performance is not sensitive to the pulse length, as long as it exceeds the cavity life-time. In fact, in the ``Pulsed Signal and Pulsed Pump'' case, a higher MER of $85.7\% W$ is seen for the same pump power than in the case with a quasi-CW signal. Also, the MER can be increased by operating in a less over-coupling regime. As of now, the power consumption for modulation is 12 pJ. By using shorter pulses closer to the cavity lifetime of $300$ ps, the modulation energy can be reducted to $\sim 1$ pJ, although we currently do not have this instrument to create such pulses. 

We would like to note that the idea of light controlling light has gained interest in the pursuit of all-optical information processing and beyond, as an indispensable element and a powerful tool. Recent demonstrations have employed two-photon absorption \cite{grinblat2019ultrafast} and second-harmonic generation and its reversion \cite{guo2022femtojoule}. In comparison, QZB uniquely allows all-optical logic in an interaction-free and excitation-free manner, promising chip-integrated optical transistor technology that is robust and scalable. For example, in the Supplemental Materials, we show how an all-optical switch is realized simply by adding a drop port. For realistic device parameters and 1 ns pulses, a 6-mW signal can be switched by a 2-mW pump with over 90\% efficiency. Meanwhile, due to interaction-free, $65\%$ of pump survives after the switch, so that it can subsequently switch another signal pulse, providing the needed cascadability for all-optical logic networks.   

Another interesting feature with QZB is that, the roles of control and signal are determined by which arrives at the cavity earlier. This gives rise to a potentially powerful operation where the conditional logic is reversed by controlling the relative time delay between the input pulses. For detailed result figures please see the Supplemental Materials.

With the above features, the QZB devices provide fan-out for all-optical logic operations. For example, two QZB switches can be cascaded to implement a complex operation, with the signal output in one to act as the input pump for the other, as long as the (quasi-)phase matching condition is still met. To this end, one may apply a frequency converter to shift the signal wavelength, or use a microcavity with multiple phase matching peaks.    

Lastly, the interaction-free and excitation-free features make QZB particularly suitable for quantum applications. Indeed, the same device shall readily work for quantum signals at a single-photon level, except for potential Raman scattering which can be suppressed by further detuning the pump and signal. Furthermore, by improving the intrinsic Q-factor to $10^8$, as demonstrated in a laser-written and chemical and mechanical polished microring \cite{gao2022lithium}, the required pump power can be reduced to single-photon level as well. This opens door to single-photon anharmonicity and quantum C-NOT gates between single photons.


\begin{acknowledgments}
The research was supported in part by the Office of Naval Research (Award No. N00014-21-1-2898). Device fabrication was performed at Advanced Science Research Center, City University of New York.
\end{acknowledgments}

\nocite{*}
\bibliography{sorsamp}

\end{document}